\documentclass[showpacs,floatfix,showkeys,preprint]{revtex4-1} 
\usepackage{amsmath,subfigure,dcolumn}  
\usepackage{amsmath,bm,esint,times,verbatim,hhline,amssymb,tipa,makeidx,hyperref,dcolumn,placeins}
\usepackage[pdftex]{graphicx}
\usepackage{amsfonts} 
\usepackage{color,url,setspace,natbib,float,rotating}
\usepackage[utf8]{inputenc} 
\usepackage{caption,wasysym}

\begin{document}

\title{Non-stationary Dynamics in the Bouncing Ball: A Wavelet perspective}

\author{Abhinna K. Behera}
\email{abhinna@iiserkol.ac.in} 
\altaffiliation[Address: ]{Department of Physical Sciences, Indian Institute of Science Education and Research (IISER) Kolkata, Mohanpur - 741252, India.} 
\affiliation{Department of Physical Sciences, Indian Institute of Science Education and Research (IISER) Kolkata, Mohanpur - 741252, India.}

\author{A. N. Sekar Iyengar}
\email{ansekar.iyengar@saha.ac.in}
\affiliation{Plasma Physics Division, Saha Institute of Nuclear Physics (SINP), Sector 1, Block - AF, Bidhannagar, Kolkata-700064, India.}

\author{Prasanta K. Panigrahi}
\email{pprasanta@iiserkol.ac.in}
\affiliation{Department of Physical Sciences, Indian Institute of Science Education and Research (IISER) Kolkata, Mohanpur - 741252, India.}

\begin{abstract}
The non-stationary dynamics of a bouncing ball, comprising of both periodic as well as chaotic behavior, is studied through wavelet transform. The multi-scale characterization of the time series displays clear signature of self-similarity, complex scaling behavior and periodicity. Self-similar behavior is quantified by the generalized Hurst exponent, obtained through both wavelet based multi-fractal detrended fluctuation analysis and Fourier methods. The scale dependent variable window size of the wavelets aptly captures both the transients and non-stationary periodic behavior, including the phase synchronization of different modes. The optimal time-frequency localization of the continuous Morlet wavelet is found to delineate the scales corresponding to neutral turbulence, viscous dissipation regions and different time varying periodic modulations. 
\end{abstract}



\maketitle

\section{Introduction} 
Driven non-linear systems are well known to exhibit complex dynamics, ranging from stable to unstable periodic motions and chaotic behavior, under different conditions. A number of simple systems e.g., coupled pendulum, electrical circuits can reveal extremely complex dynamics.$^{1, 2}$ Well known systems such as bilinear oscillator,$^{2}$ non-linear Schrödinger equation,$^{3-6}$ billiards,$^{8-9}$ Chua circuit,$^{10-11}$ have been studied quite extensively in the literature. Highly complex systems like impact oscillators, reveal complex dynamical behavior, as certain periodic orbit of the system goes through grazing inside the impacting regime.$^{7}$ Bifurcations in the periodic orbit of classical dynamical systems e.g., quartic oscillator, have been shown to have effect on the quantal dynamics.$^{12}$ All these make exploration of classical non-linear systems, an exciting area of ongoing research.

The non-linearity in these systems can arise from non-linear interactions or from imposed boundary conditions. In linear cases, the driving force can induce modulations, commensurate with the driving frequency, whereas in the case of non-linearity, other frequencies can manifest. Here, we explore the rich dynamics of the driven ``Bouncing ball" system and characterize its complex temporal behavior through Fourier and wavelet based approaches. This system has been studied earlier due to its interesting dynamics. The period doubling route to chaos in such systems in the presence of dissipation has been illustrated.$^{13-14}$ Heck et al., used high speed camera for observing the dynamics of this system, and this was then modeled numerically.$^{15}$ A computer controlled system has been configured  in order to ascertain the accurate state of the bouncing ball system.$^{16}$ Okninski and Radziszewski, have studied numerically, as well as analytically, the dynamics of a bouncing ball moving in a gravitational field and colliding with a vertically moving table with constant velocity.$^{17}$

This driven system shows non-stationary behavior and unstable periodic orbits. It also exhibits chaotic dynamics under appropriate parametric conditions. Hence, conventional Fourier based approach is inadequate to compute the local dynamics. We make use of both continuous and discrete wavelets to characterize the periodic, as well as self similar behavior of this system. The periodic components are ascertained by isolating the variations in both temporal and frequency domains. Morlet wavelet, a product of a variable Gaussian window and a sinusoidal function, is used to identify local periodic modulations. Wavelet transform splits the input time series into components that depend on both position and scale.$^{18}$ One then characterizes the variations in it, by changing the scale for a particular location. The study of power in the scale dependent variations is carried out to reveal presence of turbulence and dissipation in the system.$^{19-20}$ The multi-fractal scaling properties has also been estimated precisely through wavelet based detrended fluctuation analysis.$^{23-24}$

The paper is organized as follows. In the ensuing section, we briefly outline the set up of the bouncing ball experiment and explicate the effect of different controlling parameters on dynamics. Sec.II is devoted to the study of the complex dynamics of this rather simple system, starting with the estimation of the Lyapunov exponents. Fourier methods are used to ascertain both periodic and self-similar nature of the dynamics. Filtering out the high frequency components is shown to reveal clearly the periodic dynamics in the phase-space. Wavelet based approach is then employed to estimate multi-fractality and study the time varying dynamics. The continuous Morelet wavelet revealed the multiple non-stationary periodic modulations and their phase dynamics in the time domain. In Sec. III, the wavelet power analysis clearly identified the parameter domains, wherein the dynamics revealed neutral turbulence and viscous dissipation.$^{21-22}$ We then conclude in Sec. IV, summarizing the results and pointing out directions for future work. 

\subsection{Materials and Methods}
\begin{figure}[h!]
	\subfigure[Schematic illustration of the experimental setup.]
	{
		\includegraphics[scale=0.25]{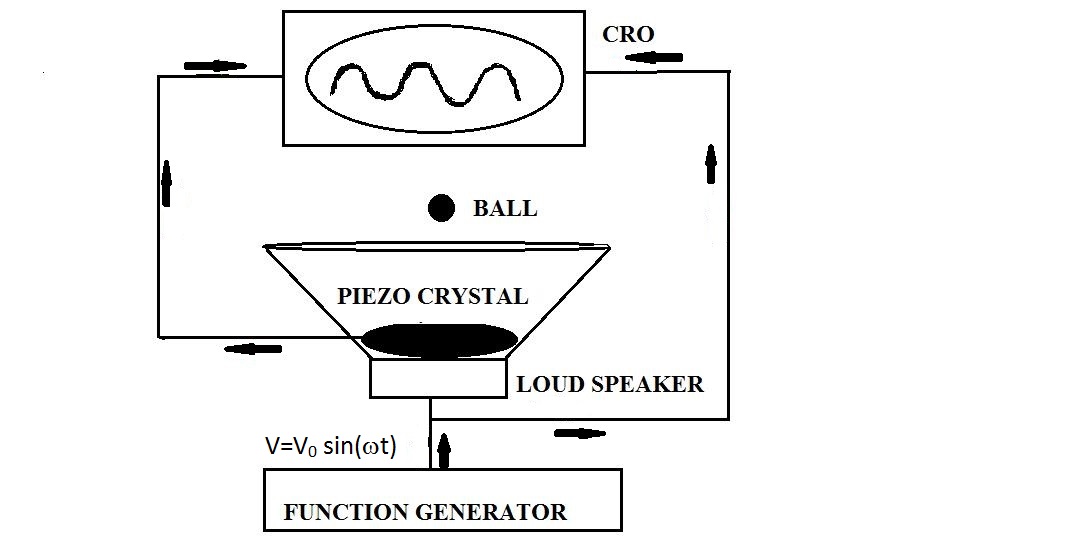}
		\label{fig:system}
	}
	\subfigure[A typical potential time series of the bouncing ball, with the corresponding denoised one shown below it, for forcing parameter values $V_{0}=40 V$ and $\omega=70 Hz$.]
	{
		\includegraphics[scale=0.17]{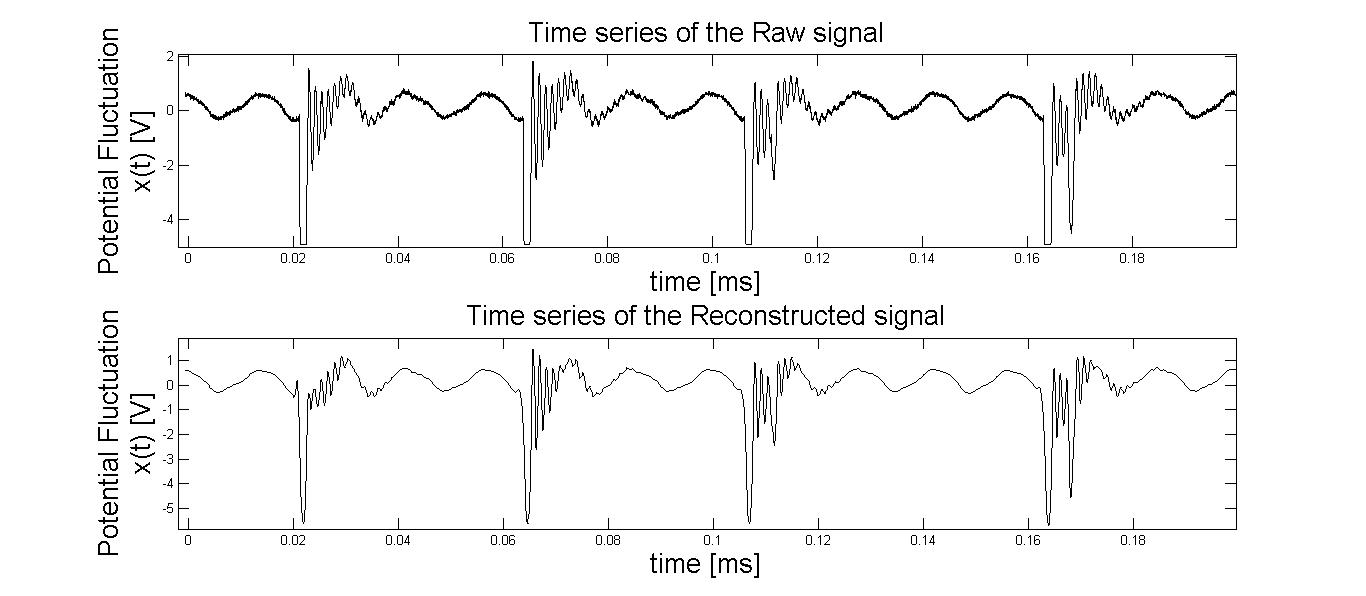}
		\label{fig:filt}		
	}
	\caption{Method for data recording, shown with a typical recorded data.}
\end{figure}

\textbf{Set up:} A loud speaker was connected to a function generator, a cathode ray oscilloscope and a resistor. A piezoelectric film was set properly on the base of the speaker to prepare the platform for the ball to bounce.$^{14, 25}$ Having fixed the input frequency of the function generator at 49 Hz, it was made to generate a sinusoidal potential, $V=V_{0} sin(\omega t)$, with $V_{0}$ and $\omega$ as the controlling parameters. Figure \ref{fig:system} depicts the experimental set up and Fig. \ref{fig:filt} shows a typical potential time series. The `Denoised' version, for highlighting the periodic motion of the raw signal, is shown for comparison. For each input voltage $V_{0}$, the frequency input $\omega$ is varied from 25 Hz to 100 Hz, in a step size of 15 Hz in order to estimate the frequency response of this system. This process was repeated for 10 V, 20 V, 30 V and 40 V input voltages. The potential fluctuations of the piezoelectric crystal $x(t)$, due to the bouncing of the ball on top of it, is the time series to be analyzed in the following sections.

\FloatBarrier

As will be seen in the following sections, the bouncing ball shows periodic behavior for smaller values of the forcing parameters. It also exhibits short-time periodic behavior and time varying transient dynamics for higher values of the potential, the unstable alteration of phases of apparently periodic time series highlights presence of intermittency.$^{25}$ As the forcing frequency increases with constant forcing voltage, the ball reveals non-linear behavior.$^{16}$ We now proceed for a quantitative study of this complex dynamics.

\section{Analysis of the Bouncing Ball Motion}
\subsection{Lyapunov Exponent}
\begin{figure}[h!]
	\subfigure[Variations in the Lyapunov exponent as a function of forcing potential, for different frequency values.]
	{
		\includegraphics[scale=0.3]{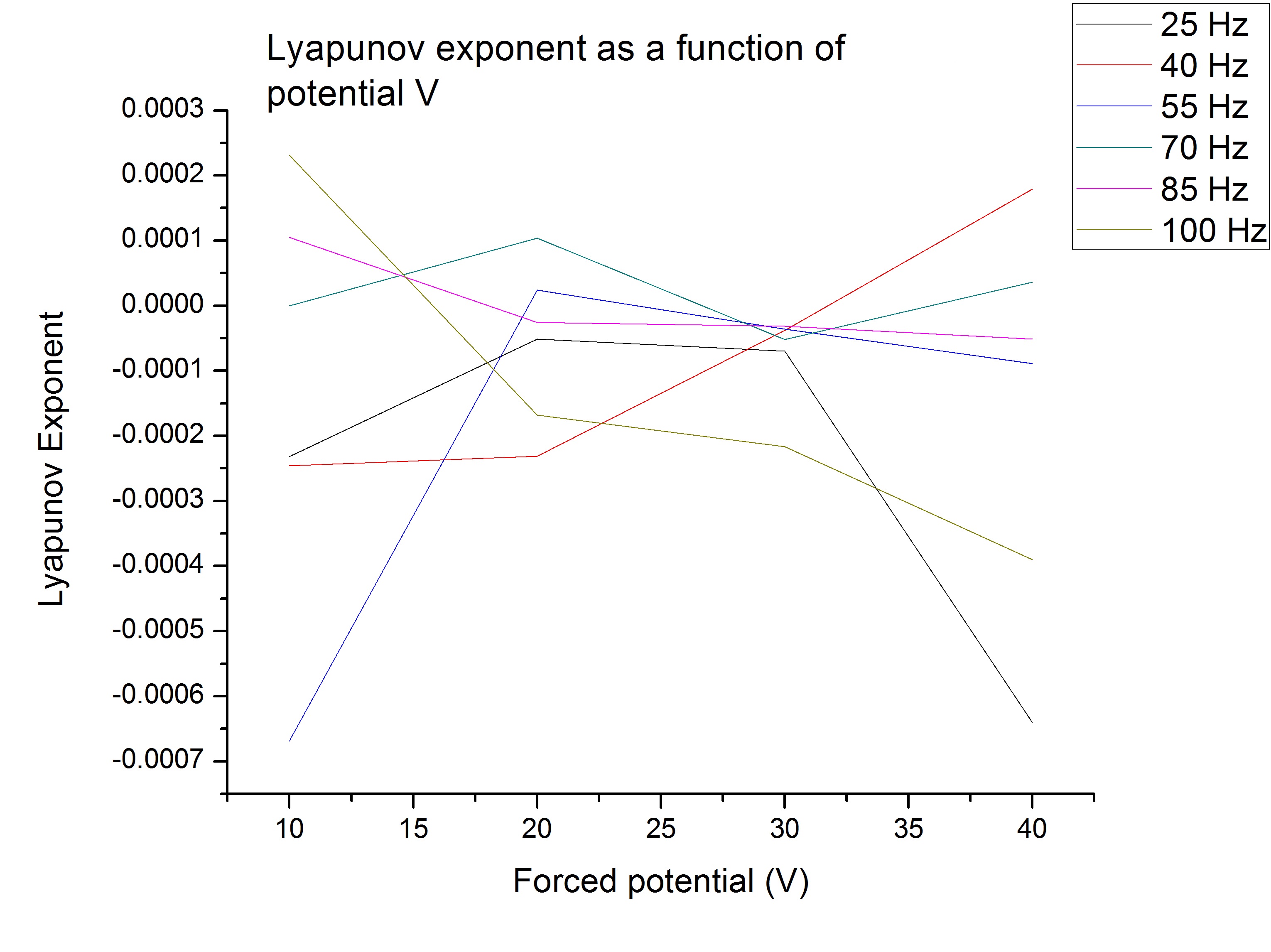}
		\label{fig:lya} 
			
	}
	\subfigure[Variations in the Lyapunov exponent as a function of forcing frequency, for different input potentials.]
	{
		\includegraphics[scale=0.3]{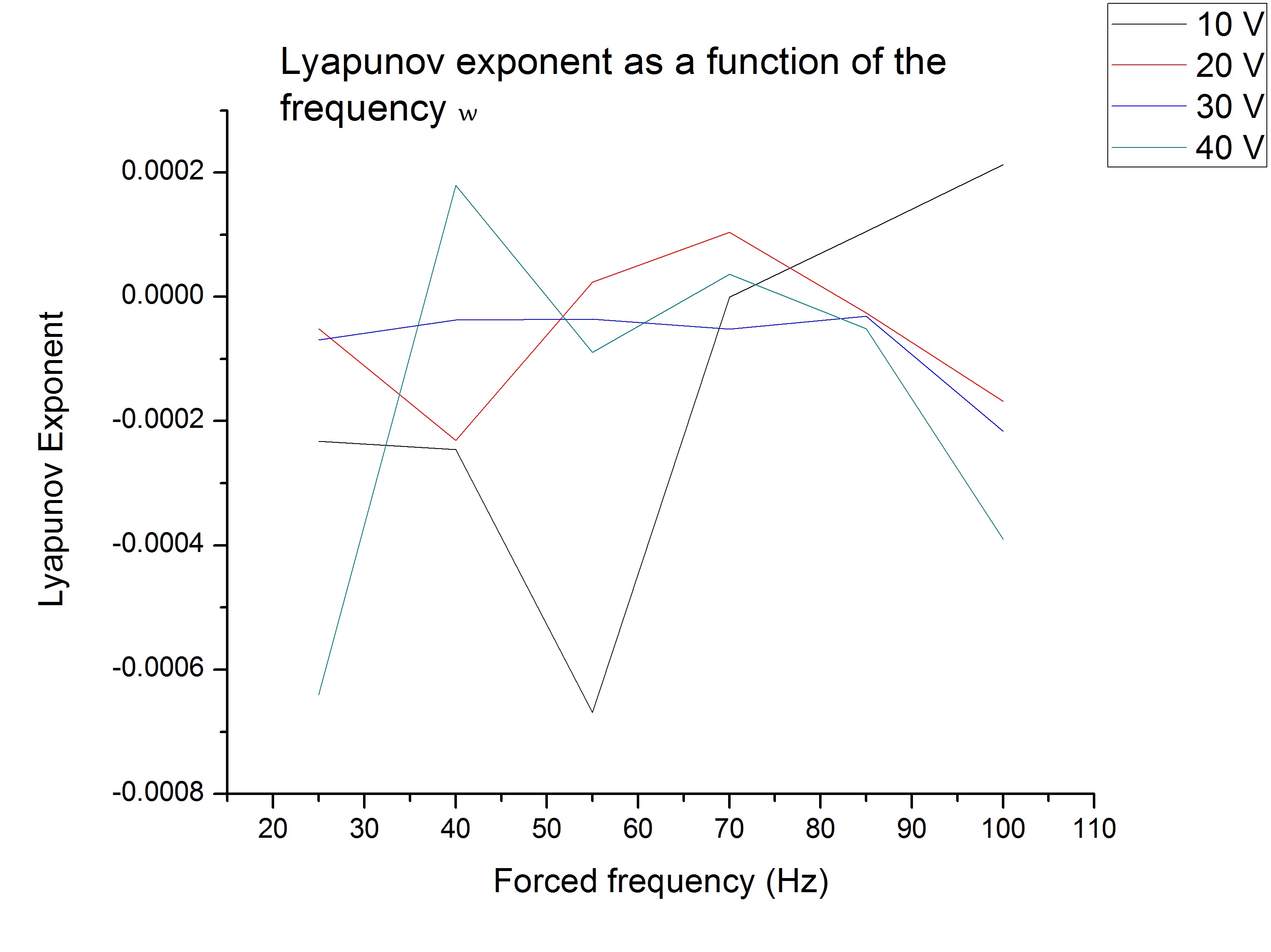}
		\label{fig:lyu} 
	}
	\caption{Variation of Lyapunov exponent, shown for different forcing parameters. }
\end{figure}

We start with the estimation of the Lyapunov exponent $(\lambda)$. As is evident from Figs. \ref{fig:lya} and \ref{fig:lyu}, for low values of the forcing parameters, the exponent is negative. Negative Lyapunov exponents are characteristic of dissipative systems, which exhibit asymptotic stability. Hence, for negative $\lambda$, the orbit is attracted to a stable fixed point; the more negative the exponent, the greater being the stability. For higher values of the forcing parameters, $\lambda > 0$ is obtained, where the system shows chaotic dynamics.$^{1}$ 

\FloatBarrier

Figures \ref{fig:lya} and \ref{fig:lyu} depict the Lyapunov exponents, obtained by varying both the forcing parameters. We note that, even for low potential input, at higher frequencies the system's $\lambda$ is positive. It is observed that, for high values of forcing potential, with low values of forcing frequency, the system can show chaotic behavior; whereas with high values of forcing frequency, the system exhibits periodic nature, though it is forced with high values of potential, except at 10 V forcing potential, which exhibits contrasting non-linear behavior. In the following, we explore in more detail, the dynamics of this potential time series, using Fourier method and Wavelet based approach.

\subsection{Frequency Domain Fourier Analysis}
Fourier transform is a well known method, which decomposes a signal to complex exponential functions of different frequencies:

\begin{equation}
X(\omega)=\int_{-\infty}^{\infty} x(t)e^{- i\omega t}dt 
\end{equation}
with the inverse transform,
\begin{equation}
x(t)=\frac{1}{2\pi} \int_{-\infty}^{\infty} X(\omega)e^{ i\omega t}d\omega ,
\end{equation}
here $\omega$ and $t$ are frequency and time variables.

\begin{figure}[h!] 
	\subfigure[The Fourier domain power spectrum of a typical signal.]
	{
		\includegraphics[scale=0.15]{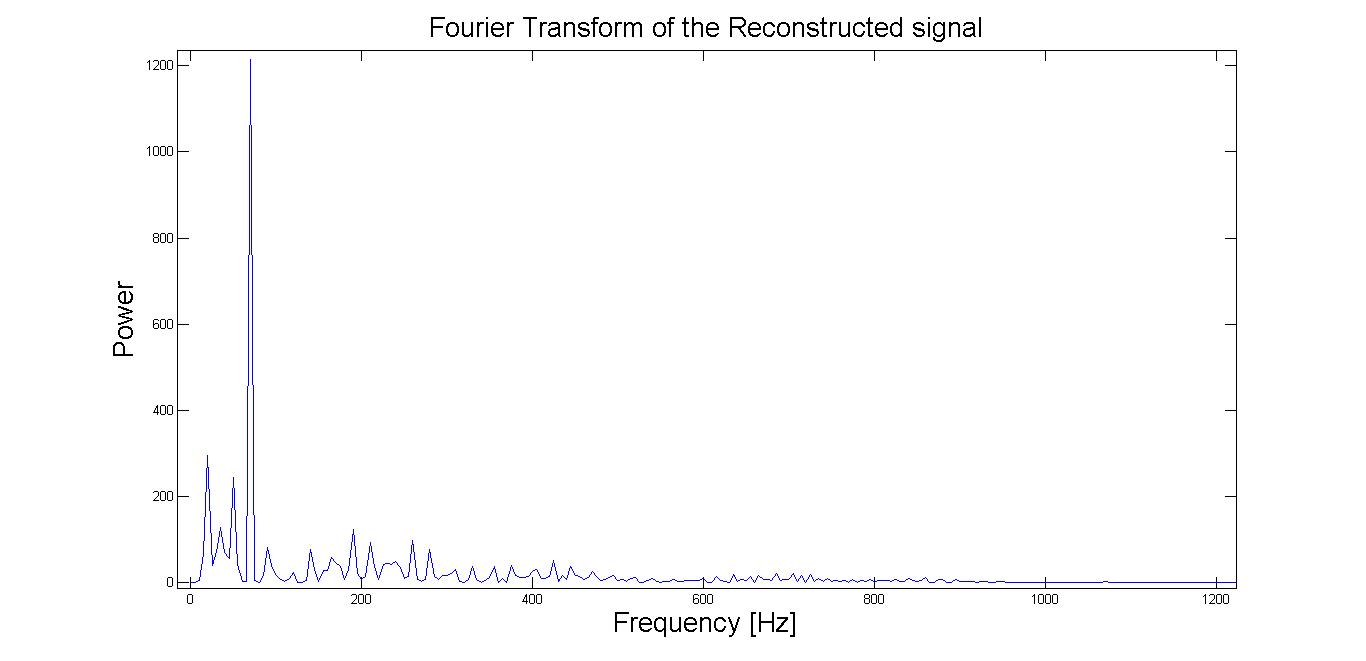}
		\label{fig:fou}
	}
	\subfigure[Phase-space plot corresponding to the `Denoised' version of the signal, clearly showing periodic motion.]
	{
		\includegraphics[scale=0.15]{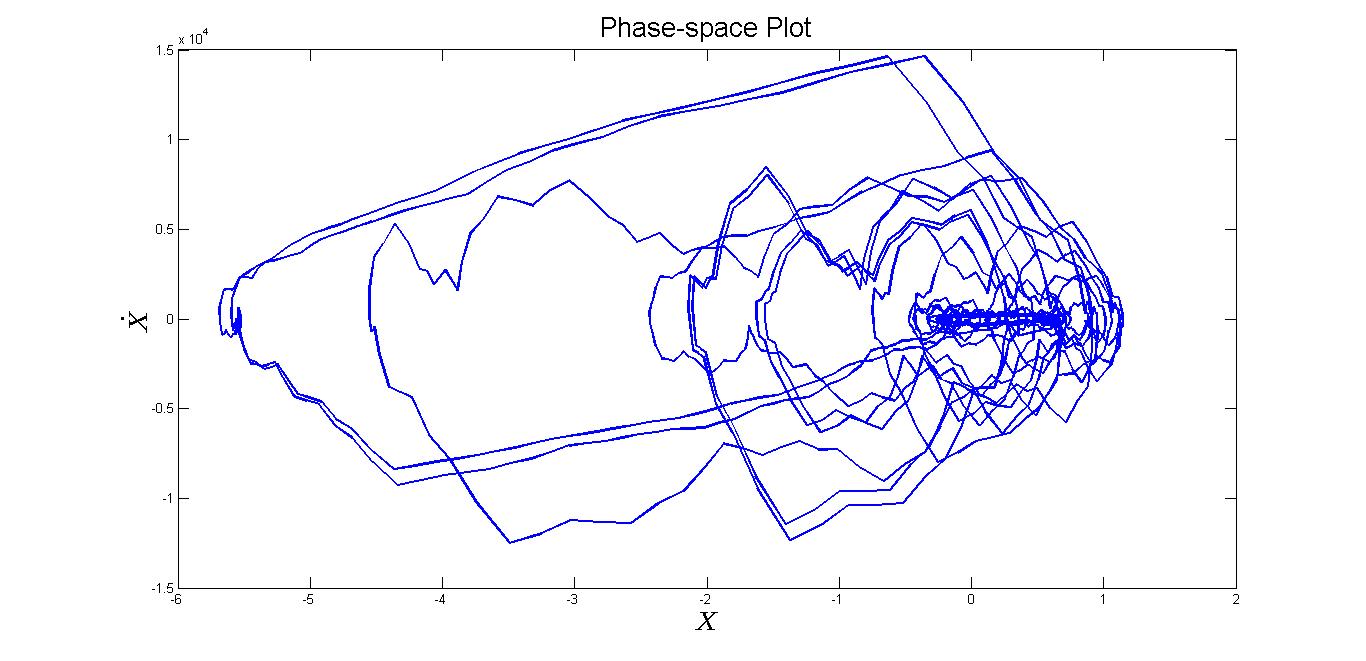}
		\label{fig:phase}
	}
	\caption{Frequency response of the system, shown with its multi-periodic behavior in the phase-space.}
\end{figure}
\FloatBarrier

Figure \ref{fig:fou} shows a typical Fourier power spectrum, showing the dominant low-frequency periodic components after denoising. As expected, the signals contain a noise component, which can be reduced through denoising to highlight the periodic behavior. The noise strength was found to increase with increasing magnitude of the forcing parameters. We have applied a wavelet based filter on the signal to remove high frequency transients to clearly highlight the periodic behavior in the phase-space.$^{7}$ Figure \ref{fig:phase} depicts the phase-space plot of the filtered signal, explicating the periodic dynamics. The number of converging loops show the multi-periodic nature of the dynamics. In Fig. \ref{fig:phase}, the random phase change of the periodic orbits, indicates the intermittency behavior of the system.

\begin{figure}[h!]
	\subfigure[Maximum output frequencies, shown for each input voltage to the system. Non-linear frequency response for 10 V input voltage is clearly seen.]
	{
		\includegraphics[scale=0.25]{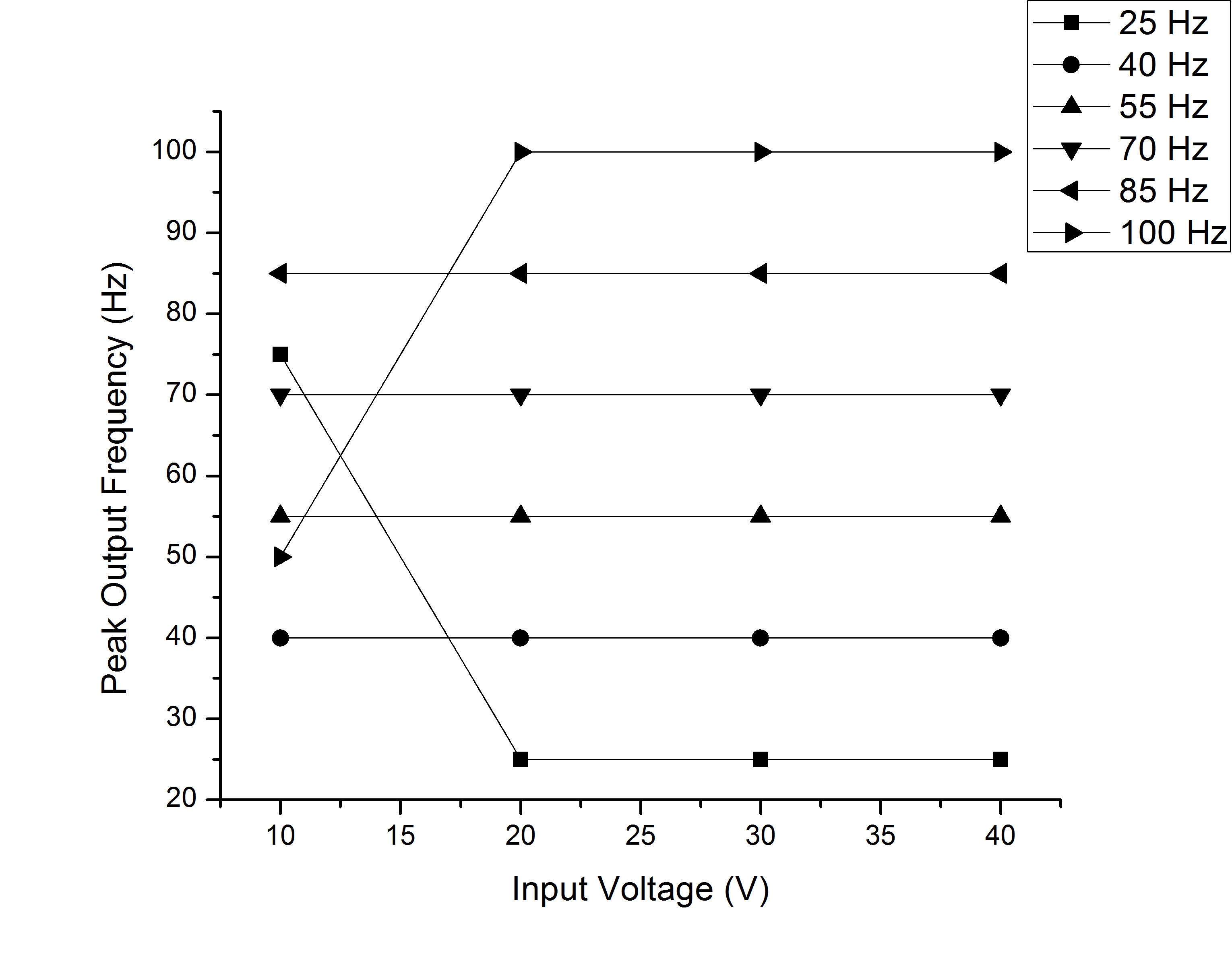}
		\label{fig:ofr}	
	}
	\subfigure[Maximum output frequencies for different input frequencies, showing non-linear behavior for frequency values 25 and 100 Hz.]
	{
		\includegraphics[scale=0.25]{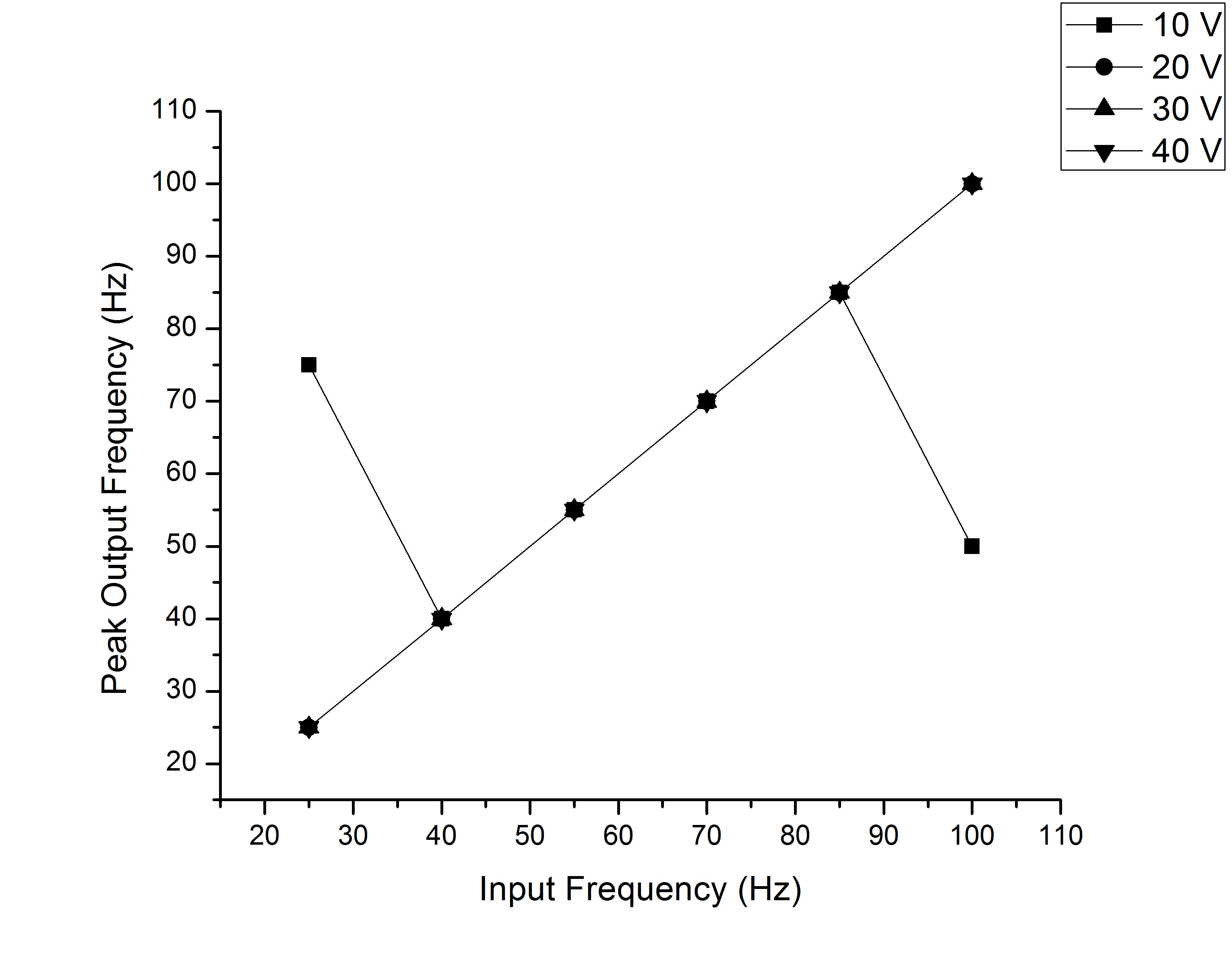}
		\label{fig:ovol}
	}
	\caption{Characteristic response frequencies of the system, shown for both the forcing parameters.}
\end{figure}
\FloatBarrier

We now study the output from the system in order to identify the parameter domain, corresponding to non-linear behavior. Figures \ref{fig:ofr} and \ref{fig:ovol} represent the response of the system to varying forcing parameters. In Fig. \ref{fig:ofr}, except for the input parameter of 10 V, the highest response frequencies are same as the respective forcing frequencies, indicating linear behavior. For $V_{0}=10V$, the highest response frequency output is found to be 75 Hz and 50 Hz, for 25 Hz and 100 Hz inputs respectively. In Fig. \ref{fig:ovol}, for higher values of forcing parameters, non-linear behavior is observed from the frequency response. The frequency response gets diminished in the system and can also get amplified for smaller values of forcing parameters.$^{16}$ This behavior, arising due to turbulence, is studied in detail in Sec.II(c).

\begin{figure}[h!]
	\subfigure[Cumulative sum of the profile: $y_{i}=\sum_{j=1}^{i}x_{j}$.]
	{
		\includegraphics[scale=0.14]{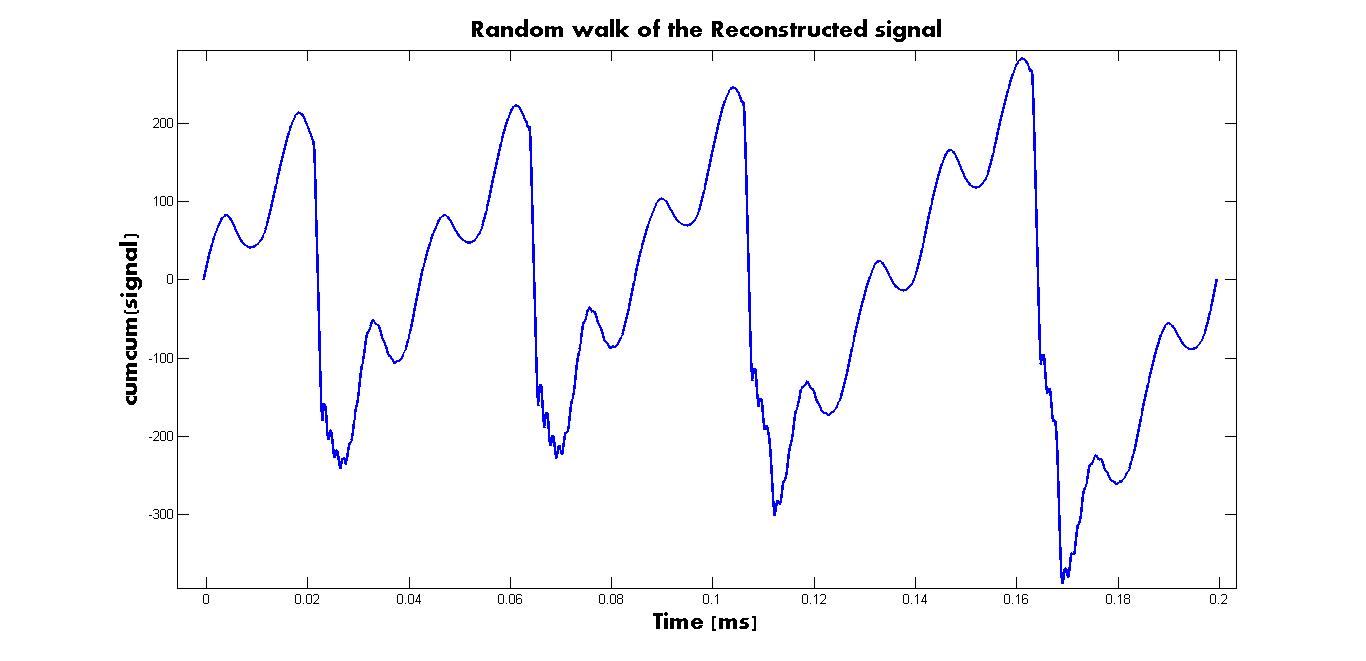}
		\label{fig:cum}	
	}
	\subfigure[Linear fit of the log-log plot of the Fourier power of the profile. In higher frequency domain, the Hurst exponent (\textit{H}) is found to be $\approx 1$, indicating extreme fractal.]
	{
	 \includegraphics[scale=0.15]{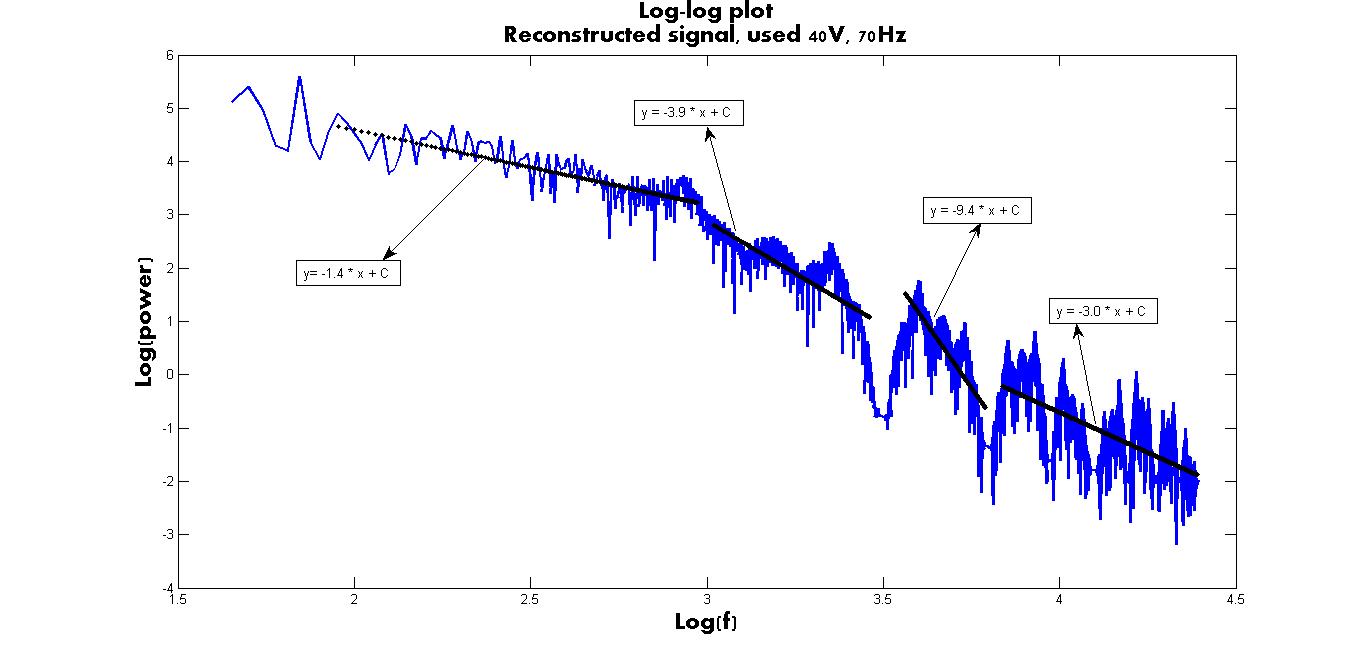}
	 \label{fig:log}
	}
\caption{The linear behavior of log-log plot of the Fourier power, showing self-similar behavior for $V_{0}=40V$ and $\omega=70Hz$. The Hurst Exponent (\textit{H}) quantifies the fractal structure.}
\end{figure}
\FloatBarrier

We now analyze the self-similar behavior of the time series, for identifying possible presence of fractal structure. For this purpose, the cumulative sum of the mean subtracted fluctuations is computed as shown in Fig. \ref{fig:cum}, revealing the fluctuation characteristics, varying from $\frac{1}{f}$ behavior to extreme fractal. Figure \ref{fig:log} depicts the log-log plot of the Fourier power of the cumulative sum of the normalized profile, which exhibits power law behavior of varying exponents, indicating the multi-fractal behavior.$^{23}$ The Hurst exponent is estimated from the Fourier power law analysis, based on the mono-fractal hypothesis. The power law exponent $\alpha$, is related to \textit{H}: $\alpha = 2H+1$.$^{23, 27-29}$ Figure \ref{fig:log} represents the linear fit of log-log plot and for mono-fractal nature, $\alpha$ is found to be $\approx-1,-3,-9$ and $-3$. It is evident that, the system has more than one exponents in the power law domain; in the high frequency range of $330Hz$ to $25000Hz$, $\frac{1}{f^{3}}$ dynamics is evident, whereas in the low frequency domain, it shows $\frac{1}{f}$ behavior.$^{30-32}$

As is well known, fractals can be subdivided in to reduced copies of the system. They are characterized by their fractal dimension \textbf{D},$^{31, 33}$ a non-integer number that quantifies the density of fractals. The number \textbf{D} measures the degree of fractal boundary fragmentation i.e., irregularity over multiple scales:

\begin{align}
D &= \begin{cases} \frac{7-\beta}{2}, \beta > 3 \\
  \frac{5-\beta}{2}, 1 < \beta < 3 \\
  \frac{3-\beta}{2}, \beta <1 \end{cases},
\end{align}
where $\beta=\vert \alpha \vert$.

In this bouncing ball system, the value of $D$ varies in between 1 - 2, depending upon the forcing potentials, as well as frequencies. As is evident, the mono-fractal behavior is not accurate; hence, we now make use of wavelet transformation for a better understanding of the local dynamics and mulit-fractal character.

\subsection{Wavelet Based Multi-fractal Detrended Fluctuation Analysis [WBMFDFA]}
Wavelet transform has both discrete and continuous nature. Discrete wavelet transform employs a series of different wavelet functions, composed of a scaling function, mother and daughter wavelets, which has a strictly finite extent.$^{33-34}$ The continuous wavelets are localized and they are not of strictly finite length.

A convolution product of the data sequence $x(t)$ with the scaled and translated version of the scaling wavelet functions $\phi(t)$ and $\psi(t)$ respectively, is known as Wavelet transform. Two parameters are used to perform scaling and translation; the scale parameter $s$ stretches (or compresses) the wavelet to the required resolution, while the translation parameter $b$ shifts the wavelet to the desired location:$^{35}$

\begin{equation}
(Wf)(s,b)=\frac{1}{s}\int_{-\infty}^{\infty}f(t)\psi^{*}(\frac{t-b}{s})dt
\end{equation}

where $s$, $b$ are real and $s > 0$. The wavelet transform acts as a mathematical microscope, with variable position and magnification, uncovering finer details of the signal, while operating at  smaller scales.$^{18}$

\subsubsection{Discrete Wavelet Transform (DWT)}
For identifying multi-fractality of fractal time series, one needs to remove the local trends. We have used discrete wavelets belonging to the Daubechies family to locate local polynomial trends from the profile. Considering $x_{i} (i = 1, 2, 3,....,N)$ to be the time series of data length $N$, the `profile' $Y(t)$ is the cumulative sum of series after subtracting the mean:
 
\begin{equation}
Y(i)=\begin{array}{c}
i\\
\sum\\
t=1\end{array}[x_{t} - <x>],\; i=1,2,3,.......N
\end{equation}

Wavelet transform on the profile $Y(t)$ has been implemented to separate the fluctuations from the trend. The fluctuation extraction have involved a wavelet decomposition using the Db4 wavelet, to remove linear trends. For Db4, in addition to the defining relation $\int \psi_{ji}dt=0$; the wavelets satisfy the vanishing moment condition $\int t\psi_{ji}dt=0$. This implies that for a linear signal of the type, $y=at+b$, the wavelet coefficients are identically zero. Therefore, this linear trend will be captured by the scaling function.$^{18, 23}$

The profile series can be decomposed for the Db4 as 

\begin{equation}
Y(t)=\sum_{i=-\infty}^{\infty}c_{i}\phi_{i}(t)+\sum_{j\geq0}\sum_{i=-\infty}^{\infty}d_{ij}\psi_{ij}(t)
\end{equation}

where, $\psi_{ij}(t)$ is the wavelet function of $Db4$ and $\phi_{i}(t)$ is the father wavelet. The coefficients $c_{i}$, $d_{ij}$ are known as the low and high pass coefficients. The profile is reconstructed at a particular level $j$, by taking only the low pass coefficients to extract the trend at level $j$. This trend is subtracted from the profile to obtain the fluctuations at that scale. Due to the convolution errors, the obtained fluctuations can have edge artifacts, which are removed by performing this fluctuation extraction on the reversed profile and taking the average. After that,  we subdivide the fluctuations in to $n_{s}$ non-overlapping segments, such that $n_{s}=\left[N/s\right]$, where the segment length $s$ is related to the wavelet scale $j$ by the number of filter coefficients used in the wavelet. The $q^{th}$ order fluctuation function, $F_{q}(s)$ is obtained as follows:$^{24}$

\begin{equation}
F_{q}(s)=\left[\frac{1}{2M_{s}}\sum_{b=1}^{2M_{s}}\left[F^{2}(b,s)\right]^{q/2}\right]^{1/q},
\end{equation}

Here $q$ can take both positive and negative integral values. If the time series  possesses fractal behavior, then $F_{q}(s)$ exhibits a power-law scaling,

\begin{equation}
F_{q}(s)\sim s^{h(q)},
\end{equation}

For $q < 0$, $h(q)$, the generalized Hurst exponent, captures the scaling properties of the small fluctuations, whereas for $q>0$, it captures the large fluctuations. $q = 0$, leads to divergence of the scaling exponent, for which logarithmic averaging has to be used to find the fluctuation function,

\begin{equation}
F_{0}(s)=exp\left[\frac{1}{2M_{s}}\sum_{b=1}^{2M_{s}}ln\left[F^{2}(b,s)\right]^{q/2}\right]^{1/q}
\end{equation}

For mono-fractals $h(q)$ values are independent of $q$, whereas for multi-fractal time series, $h(q)$ depends on $q$. The correlation behavior is characterized from the Hurst exponent $(H = h(q = 2))$, which varies from $0 < H <  1$.$^{23-24}$

Table \ref{tab:PPer} shows the Hurst exponents obtained through DWT for different forcing parameters.

\begin{table}[h!]
\caption{Hurst exponents for the potential time series with varying forcing parameters.}
\centering         
\begin{tabular}{ccccc|}    
\toprule     
Frequency $(\omega)$  & \multicolumn {3}{c}{Potential ($V_{0}$)} \\ [0.05ex]
\hline   
& 20 & 30 & 40 \\[0.05ex]
\hline
25  & 0.0543 & 0.1090 & 0.5153 \\[0.05ex] 
40  & 0.2609 & 0.5429 & 0.6644 \\[0.05ex] 
55  & 0.5999 & 0.6778 & 0.7083 \\[0.05ex] 
70  & 0.6613 & 0.7116 & 0.7338 \\[0.05ex]
85  & 0.7235 & 0.7173 & 0.7334 \\[0.05ex]
100 & 0.6888 & 0.7171 & 0.7344 \\[0.05ex]
\hline
\end{tabular}
\begin{tabular}{cc}
\toprule
Frequency $(\omega)$  & Potential ($V_{0}$) \\ [0.05ex]
\hline   
& 10 \\[0.05ex]\hline
25 & 0.1981\\[0.05ex] 
40 & 0.4652\\[0.05ex] 
55 & 0.5806\\[0.05ex] 
70 & 0.6286\\[0.05ex] 
85 & 0.2860\\[0.05ex] 
100 & 0.3466\\[0.05ex] 
\hline
\end{tabular}
\label{tab:PPer}
\end{table}
\FloatBarrier

\begin{figure}[h!]
	\subfigure[Log-log plot of the fluctuation function $F_{q}(s)$ vs. scale ($s$) for various values of $q$, for the signal with $V_{0}=40V$ and $\omega=70Hz$.]
	{
		\includegraphics[scale=0.15]{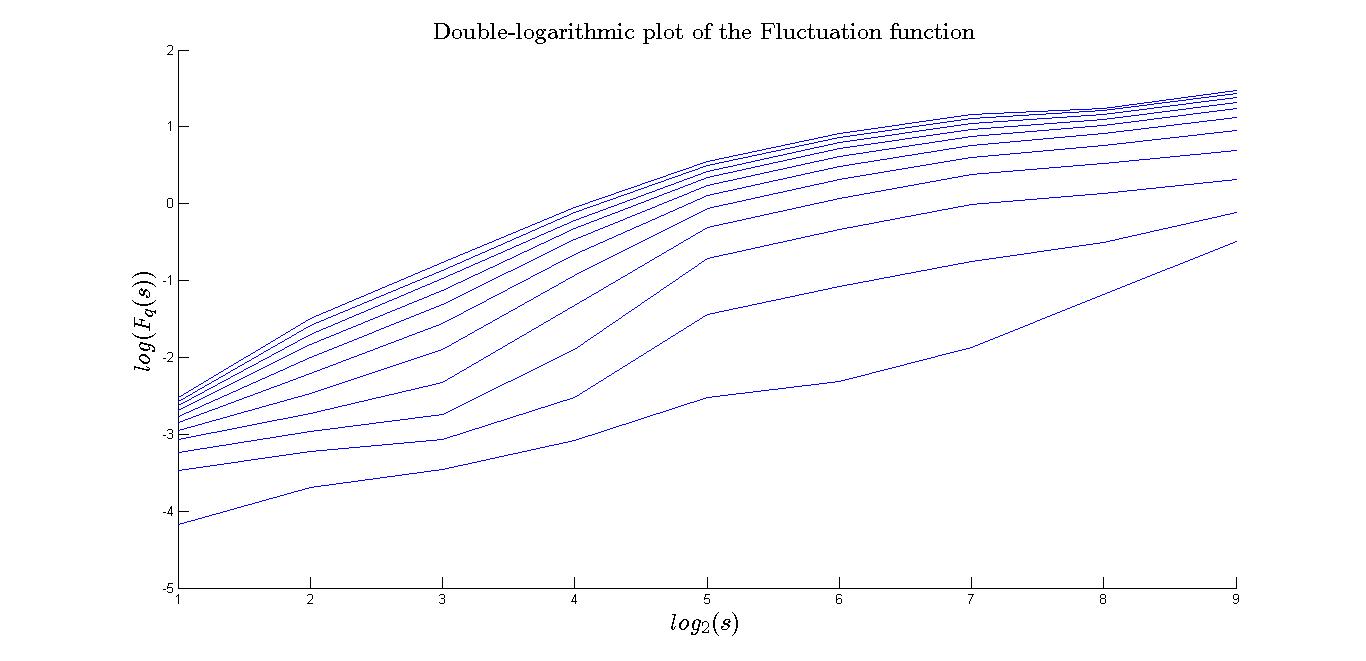}
		\label{fig:fluc}	
	}
	\subfigure[The dependence of the generalized Hurst exponent $h(q)$, as a function of the order of the moment $(q$), shown for $V_{0}=40V$ and $\omega=70Hz$, revealing the multi-fractal behavior.]
	{
		\includegraphics[scale=0.15]{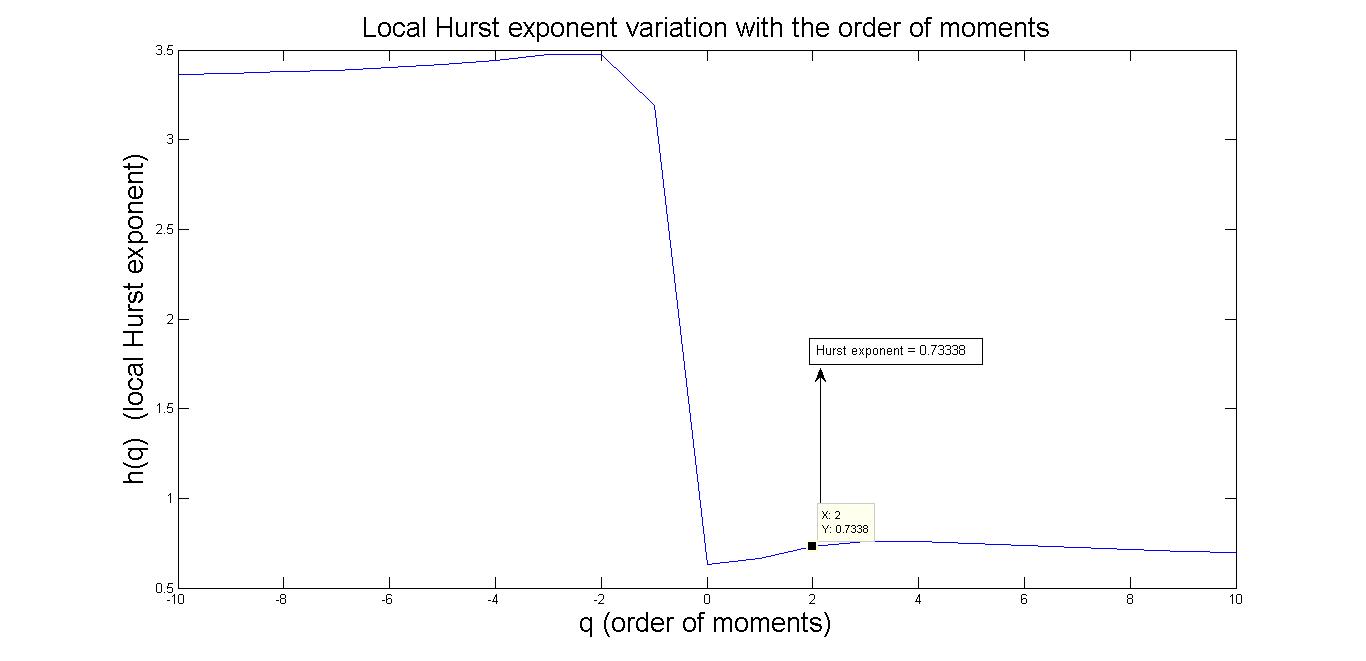} 
		\label{fig:hq}
	}
\caption{Multi-fractality of the system, shown for $V_{0}=40V$ and $\omega=70Hz$ signal.}
\end{figure}

The tabulated values explicitly shows the increase and decrease of Hurst exponents, depending on the forcing potentials and frequencies. It increases with the higher values of both the forcing parameters, except for $V_{0}=10V$. From Figs. \ref{fig:ofr} and \ref{fig:ovol}, we have already seen this contrasting feature of the system, and will explore the details of this behavior using Heisenberg and Kolmogorov fits.$^{22}$ The Hurst exponent first increases and then decreases for higher forcing frequencies. None of these multi-fractal trends could be gleaned using the Fourier analysis, where one can only qualitatively predict the overall nature of the self-similar behavior. Figures \ref{fig:fluc} and \ref{fig:hq} depict the nature of the fluctuation functions and the generalized Hurst exponents respectively. It is found that for high values of the forcing parameters, the system shows persistence, as it maintains the Hurst exponent, $H\approx 0.7$.

\FloatBarrier

Figure \ref{fig:fluc} depicts the \textit{log-log} plots of the fluctuation function $F_{q}(s)$, for $q$ between -10 to 10, in the ascending order.

We now proceed to analyze the periodic behavior through continuous wavelet transform (CWT). Both non-stationary periodic dynamics and the stationary ones clearly manifested in the scalogram of the CWT. We have used Morlet wavelet having optimal time-frequency localization. For a clearer identification of the periodic components, we removed the high frequency fluctuations through the Db4 basis. We have checked that the periodic components are not affected in this process.$^{33, 35}$

\begin{figure}[h!]
	\subfigure[$3D$ plot of the CWT coefficients with time and scale for 40 V and 70 Hz forcing parameters. Periodic behavior can be observed at certain scales.]
	{
		\includegraphics[scale=0.16]{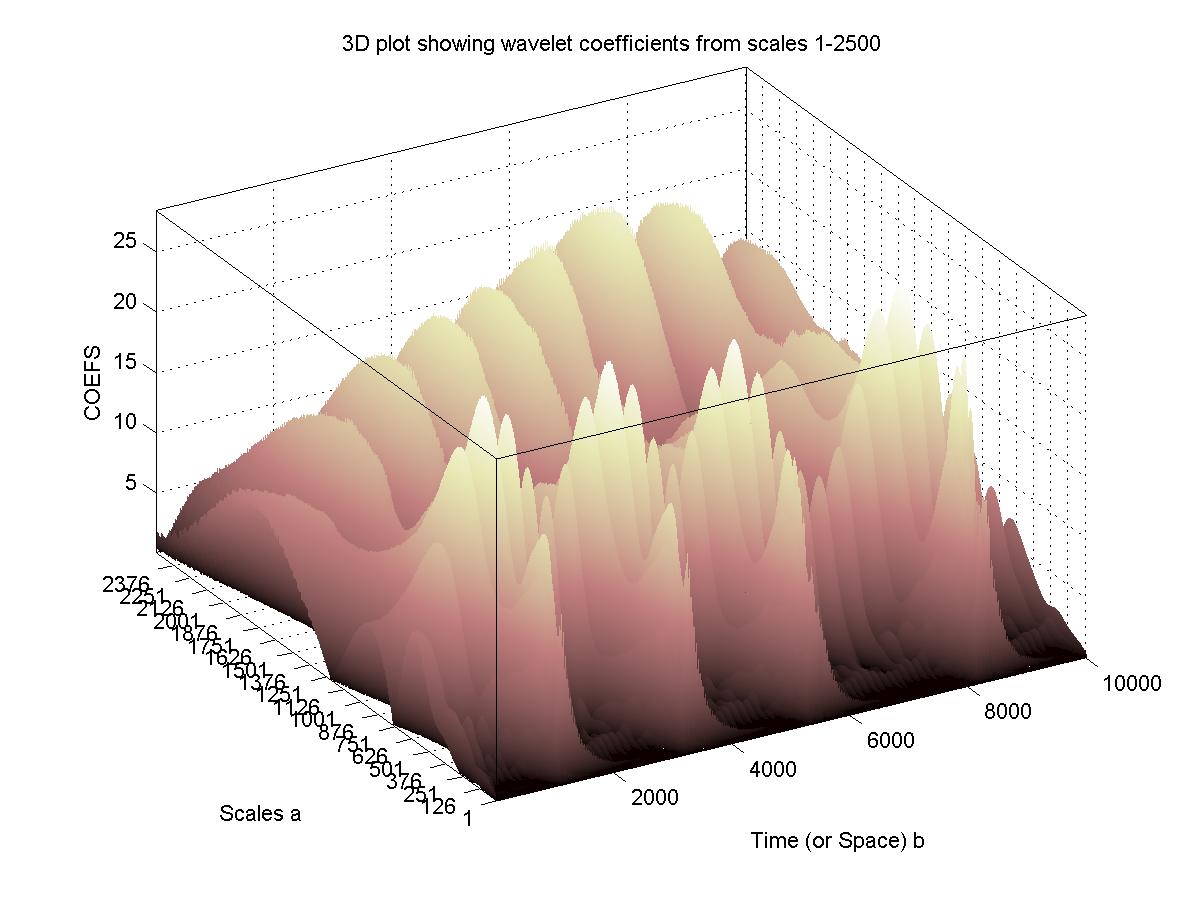}
		\label{fig:3d}	
	}
	\subfigure[Scalogram showing energy for each wavelet coefficients, with 40 V and 70 Hz as forcing parameters, clearly demonstrating the dominant periodic components.]
	{
		\includegraphics[scale=0.16]{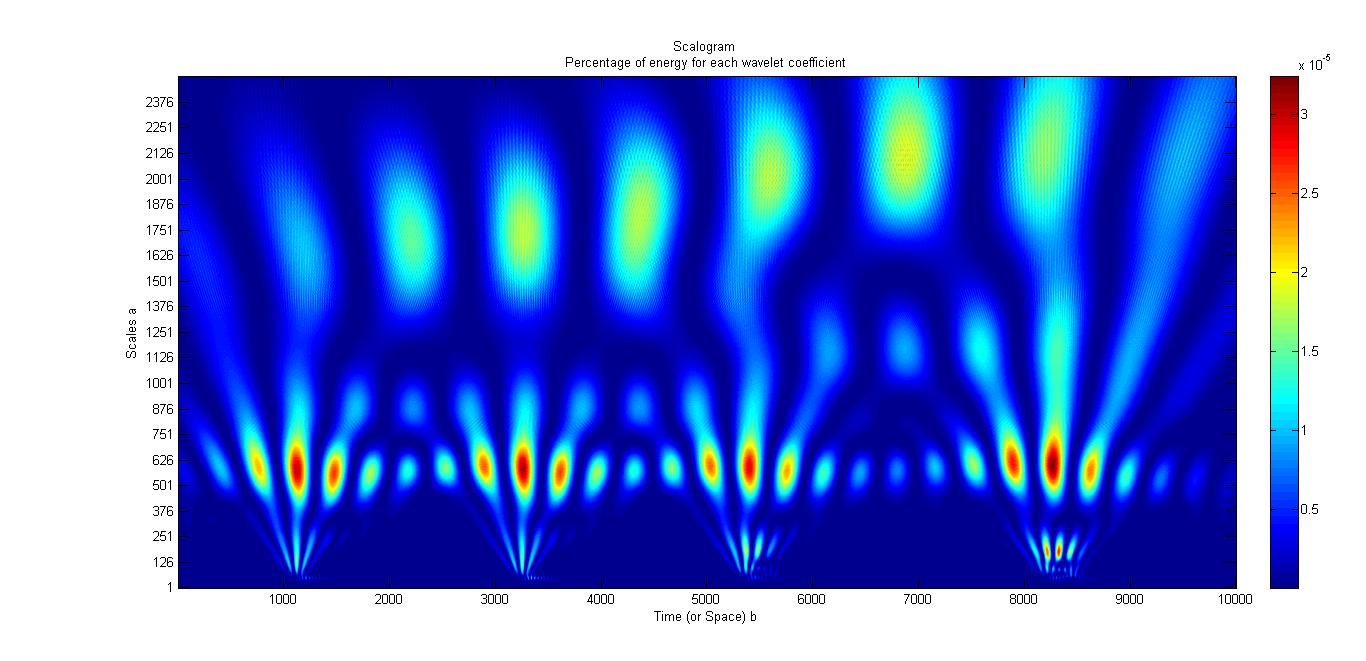}
		\label{fig:scal}
	}
\caption{CWT coefficients with its periodic nature, shown for $V_{0}=40V$ and $\omega=70Hz$ signal.}
\end{figure}
\FloatBarrier

Morlet wavelet clearly brings out different periodic components in the data sets, as seen in the scalograms given in Fig. \ref{fig:3d} in a 3D plot. Figure \ref{fig:scal} shows the scalogram of the potential time series, presenting the energy of the wavelet coefficients.$^{33, 35}$

\begin{figure}[h!]
	\subfigure[Semilog plot of wavelet power, summed over all time at different scales. It is inferred that, the ball has periodicity of 18 ms, 49 ms, 226 ms and 578 ms corresponding to 40 V and 70 Hz forcing parameters.]
	{
		\includegraphics[scale=0.16]{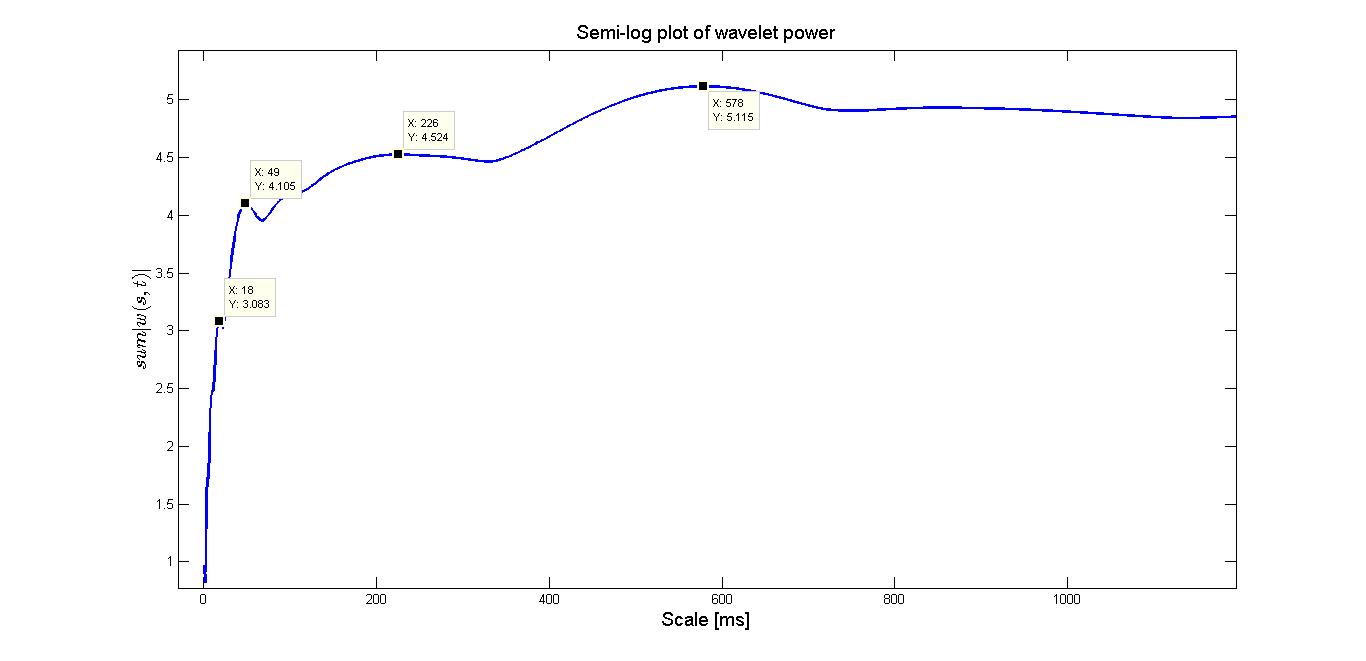}
		\label{fig:semi}
	}
	\subfigure[Wavelet power spectrum with the cone of influence. The two dominant periods of the time series can be ascertained with the confidence level of 95\%. Image colors are a representation of the Wavelet Power Spectrum (WPS) normalized by variance.]
	{
		\includegraphics[scale=0.16]{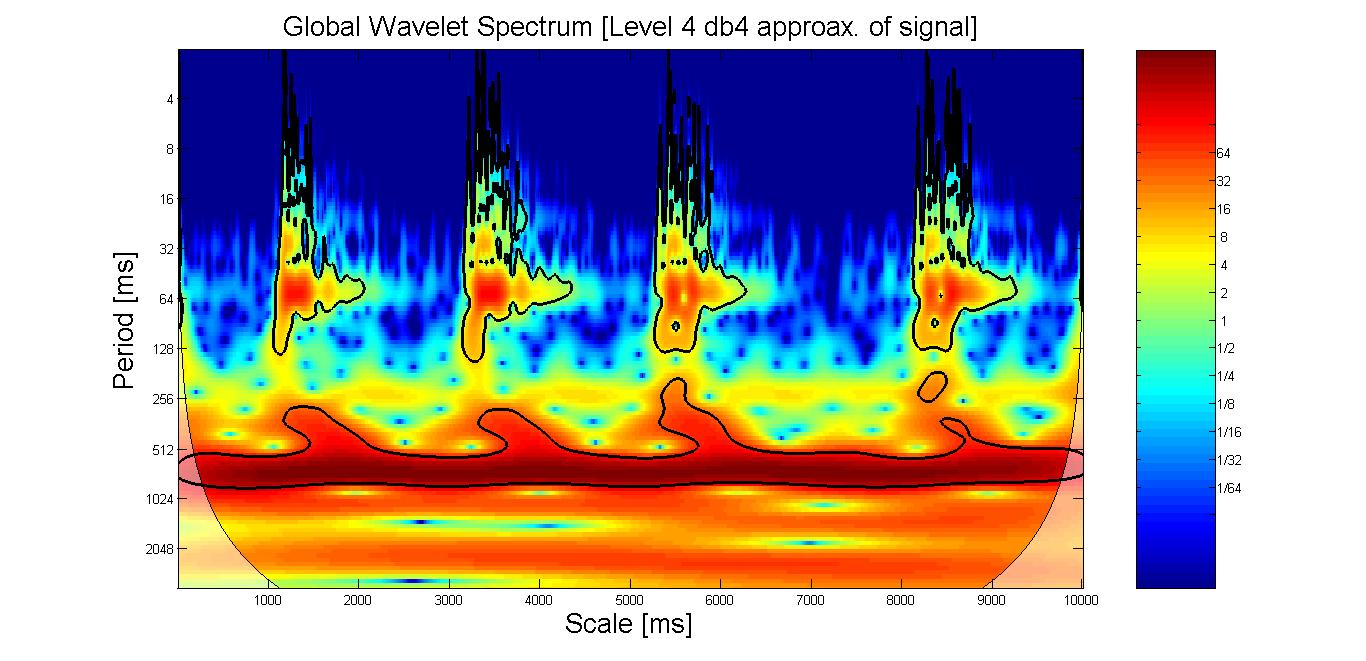}
		\label{fig:glob}
	}
\caption{Periodicity of the system, shown for $V_{0}=40V$ and $\omega=70Hz$ signal.}
\end{figure}
\FloatBarrier

It is evident that there are four dominant periods. Figure \ref{fig:glob} depicts the global wavelet power spectrum, clearly showing the presence of two periods in the potential time series. The regions of higher power (the red regions), correspond to the periods of $49ms$ and $578ms$, with 95\% confidence level.$^{33,35}$ These periods compare well with the ones obtained through Fourier transform earlier.

\begin{figure}[h!]
\centering
	\subfigure[Phase analysis with varying frequencies for $V_{0}=40V$ with $\omega=40, 55, 70$ Hz. The amplitudes of the CWT coefficients, phase of the CWT coefficients and phase difference domains, shown respectively, for  the signal of $\omega=55$ Hz in common.]
	{
		\includegraphics[scale=0.15]{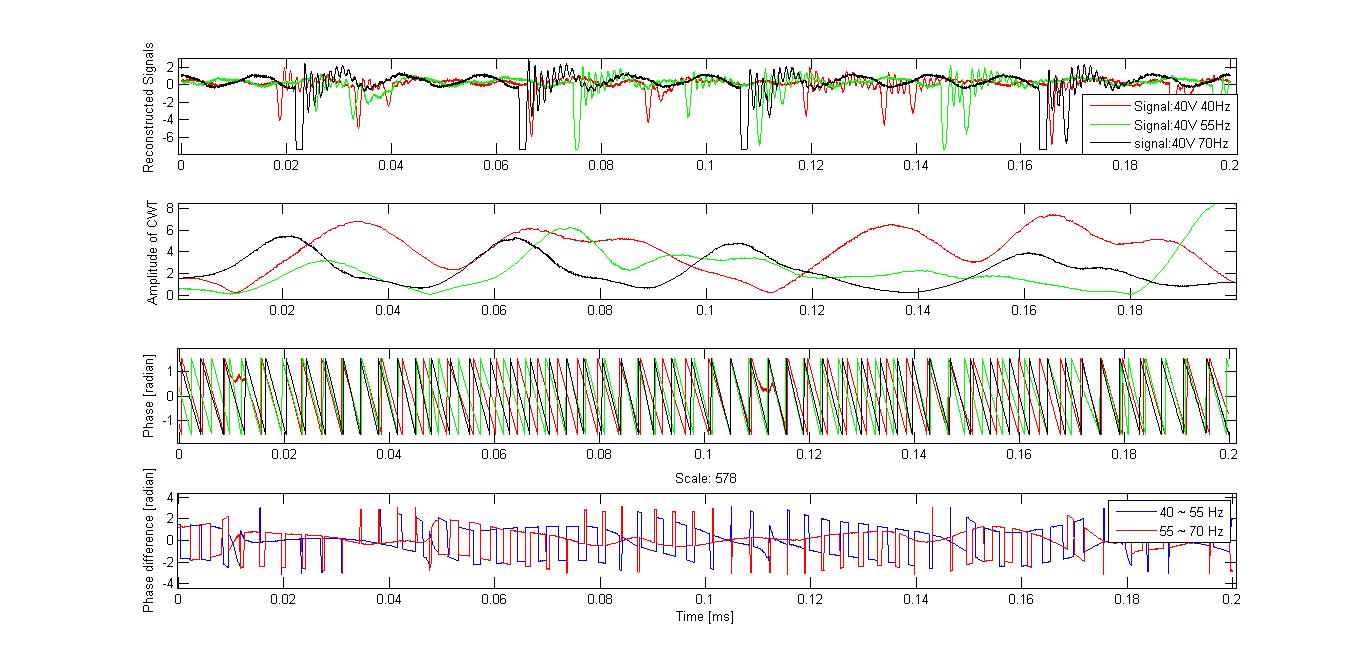}	
		\label{fig:phs1}
	}
	\subfigure[The time series considered, are the reconstructed data, for $\omega=70Hz$ having $V_{0}=20V$, $30V$ and $40V$ as varying potentials. Variation of phase and amplitude of CWT coefficients at scale 578, showing the synchronization for signal of $V_{0}=30V$.]
	{
		\includegraphics[scale=0.17]{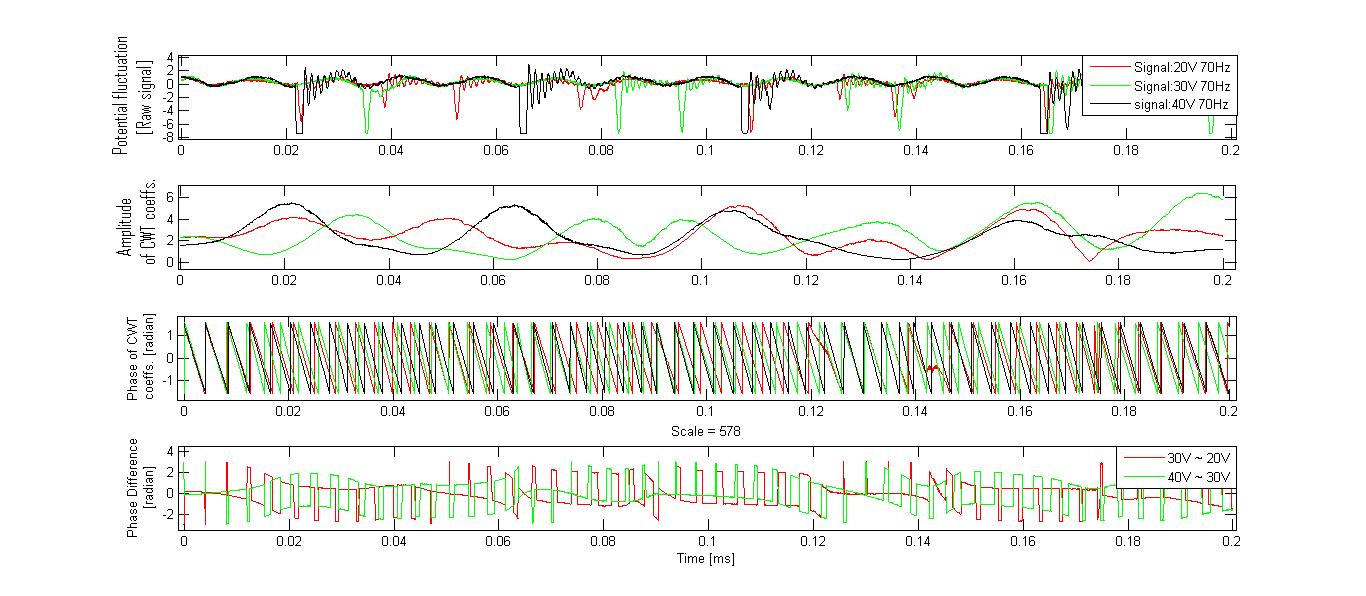}
		\label{fig:phs2}	
	}
	\caption{Phase synchronization, shown for both the forcing frequency as well as potential.}
\end{figure}
\FloatBarrier

The study of the phase structure of different periodic modulations and their possible synchronization is of great interest, as it can throw light on the underlying dynamics. For that purpose, the two dominant of periodic components are analyzed.

Figures \ref{fig:phs1} and \ref{fig:phs2}, reveal that the phase difference remains zero for certain interval of time. This behavior confirms that in certain frequency range, as observed in Fig. \ref{fig:log}, both the periodic components behave in unison. Hence, `phase synchronization' occurs in between the signals having a common forcing parameter.$^{36}$ For very small time interval, the components remain in same phase. This has a cyclic behavior, they split into a reiterating sequences of phase angles over consecutive periods.

\begin{figure}[h!]
\centering
	\subfigure[Heisenberg fits for $-\frac{5}{3}$ and $-7$ exponents, shown for $V_{0}=10V$ and $\omega=25Hz$.]
	{
		\includegraphics[scale=0.17]{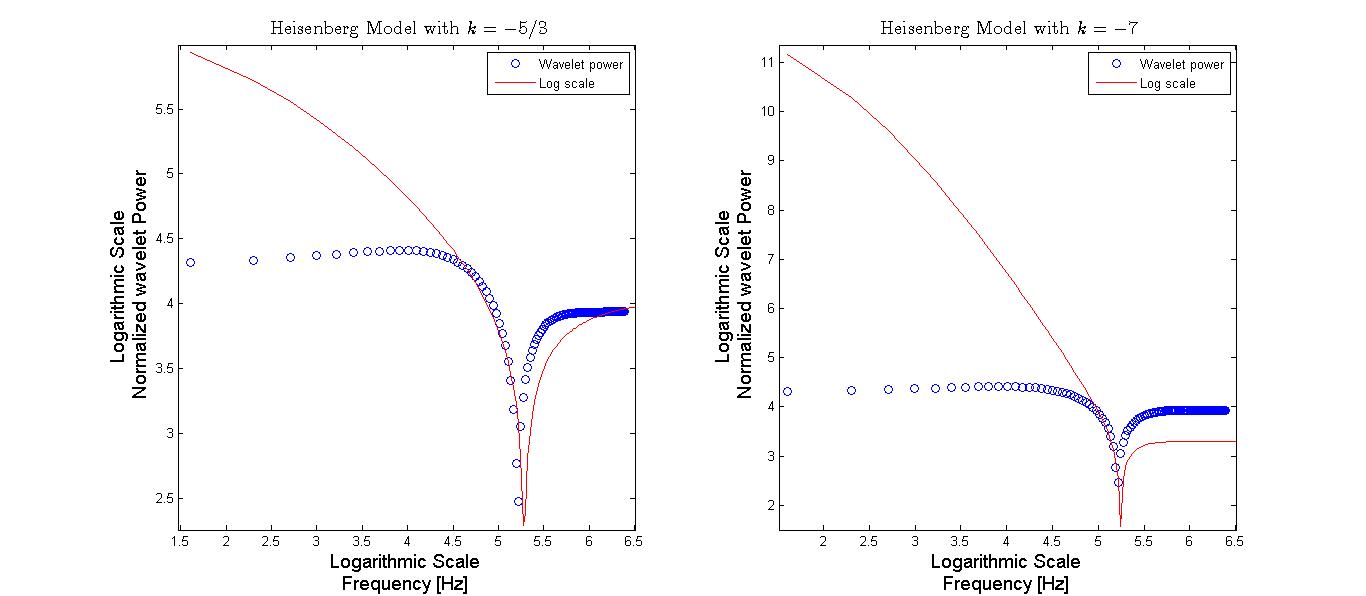}	
		\label{fig:hei2}
	}
	\subfigure[Heisenberg fits for $-\frac{5}{3}$ and $-7$ exponents, shown for $V_{0}=10V$ and $\omega=100Hz$.]
	{
		\includegraphics[scale=0.155]{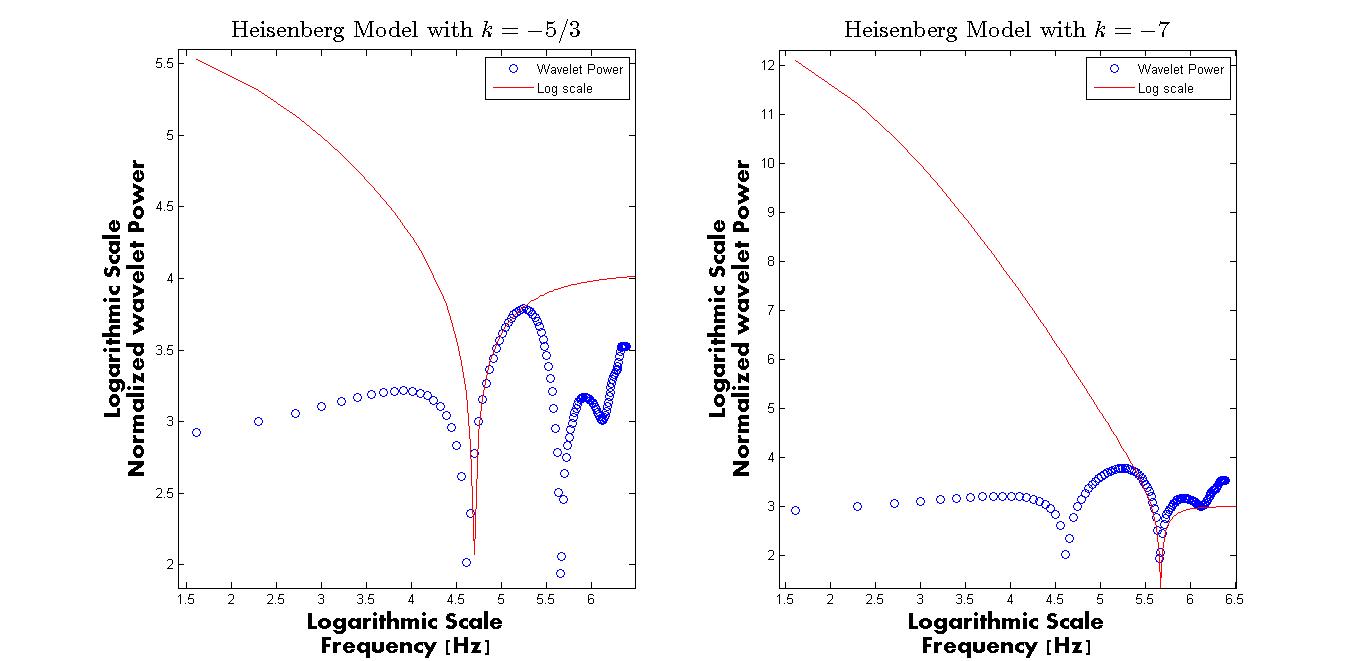}	
		\label{fig:hei3}
	}
\caption{\label{fig:heis}The normalized wavelet power spectrum Vs frequency on a log scale. The solid line is the Heisenberg fit. The first fit has a slope of $\frac{-5}{3}$, representing the neutral turbulence and the other has the slope of $-7$, exhibiting the viscous dissipation regime.}
\end{figure}
\FloatBarrier

We now explore the possible presence of turbulence as well as dissipation in the system. In Fig. \ref{fig:heis}, the wavelet power is shown, where Heisenberg fits of different exponents, $\frac{-5}{3}$ and $-7$, representing the turbulence and viscous dissipation nature of the system respectively.$^{19-20}$ As shown, some regions of the system completely fit with the curve. Hence, it is evident that the bouncing ball system has both these properties for certain parameter domain. The calculated Lyapunov exponents of the system (Fig. \ref{fig:lya}) depicts this behavior previously.$^{17}$ Figures \ref{fig:hei2} and \ref{fig:hei3} highlight the nonlinear nature of the system, as found in Figs. \ref{fig:ofr} and \ref{fig:ovol} for $\omega=25Hz$ and $\omega=100Hz$ having $V_{0}=10V$ respectively. 

\section{Conclusion}
In conclusion, we have studied the dynamics of bouncing ball, a driven system exhibiting complex dynamics, controlled by the potential and frequency of the driving source. Depending on the driving parameters, it shows self-similar and periodic behavior, which are studied using the Fourier domain analysis and Wavelet transform. It is observed that the potential time series shows a long term switching between high and low frequency, because of the high value of Hurst exponent $(H > 0.5)$. The characterization of the self-similarity revealed multi-fractal behavior. Wavelets belonging to the Daubechies family optimally removed the local trends of the time series, from which the multi-fractal character has been reliably extracted. The continuous Morlet wavelet localized the periodic components, which revealed the linear and non-linear behavior. It is evident that the bouncing ball system provides an excellent experimental tool to study nonlinear dynamics and chaos. The apparatus can be used to study the complex chaotic or periodic motions, depending on the parametric conditions of the ball.

\appendix*   

\begin{acknowledgments}
I acknowledge Mr. B. K. Giri and Mr. C. Mitra, for extensive discussions in the course of this work.
\end{acknowledgments}


\begin{thebibliography}{99}
\bibitem{key-1} S. H. Strogatz, \textit{Non-linear Dynamics and Chaos: with application to physics, biology, chemistry and engineering}, Advance Book Program, Perseus Books, Massachusetts (1994); M. Lakshmanan and S. Rajasekar, \textit{Nonlinear Dynamics: Integrability, chaos and patterns}, Springer-Verlag Berlin Heidelberg (2003) and references therein.
\bibitem{key-2}S. Banerjee, \textit{Dynamics for Engineers}, John Wiley \& Sons Ltd, The Atrium, Southern Gate, Chichester, West Sussex $PO19$ $8SQ$, England (2005) and references therein.
\bibitem{key-3}C. Sulem and P. L. Sulem, ``The Nonlinear Schrödinger Equation: Self-Focusing and Wave Collapse'', Applied Mathematical Sciences \textbf{139}, Springer-Verlag Berlin Heidelberg (1999) and references therein.
\bibitem{key-4}I. V. Barashenkov, E. V. Zemlyanaya and M. $B\ddot{a}r$, ``Travelling solitons in the parametrically driven nonlinear Schrödinger equation'', Phys. Rev. E \textbf{64}, 016603 (1-12) (2001).
\bibitem{key-5}T. S. Raju, C. N. Kumar and P. K. Panigrahi, ``On exact solitary wave solutions of the nonlinear Schrödinger equation with a source'', J. Phys. A: Math. Gen. \textbf{38} (16), L271-L276 (2005); V. M. Vyas, T. S. Raju, C. N. Kumar and P. K. Panigrahi, ``Soliton solutions of driven nonlinear Schrödinger equation'', J. Phys. A: Math. Gen. \textbf{39}, 9151-9159 (2006).
\bibitem{key-6}A. Khare and A. Saxena, ``Linear superposition for a class of nonlinear equations'', Phys. Lett. A \textbf{377} (39), 2761-2765 (2013).
\bibitem{key-7}S. Banerjee, E. Pavlovskaia, J. Ing and M. Wiercigroch, ``Complex dynamics of bilinear oscillator close to grazing'', Int. J. Bifurcation and Chaos \textbf{20} (11), 3801-3817 (2010) and references therein.
\bibitem{key-8}Y. G. Sinai, ``Dynamical systems with elastic reflections'', Russ. Math. Surv. \textbf{25} (2), 137-189 (1970) .
\bibitem{key-9}M. V. Berry, ``Quantizing a classically ergodic system: Sinai's billiard and the KKR method'', Annals of Physics \textbf{131}, 163-216 (1981).
\bibitem{key-10}L. O. Chua, M. Itoh, L. Kocarev and K. Eckert, ``Chaos synchronization in Chua's circuit'', J. Circuit Syst. Comp. \textbf{03} (1), 93-108 (1993).
\bibitem{key-11}P. Muruganandam, K. Murali and M. Lakshmanan, ``Spatiotemporal dynamics of coupled array of Murali–Lakshmanan–Chua circuits'', Int. J. Bifurcation and Chaos \textbf{09} (5), 805-830 (1999).
\bibitem{key-12}M. S. Santhanam, V. B. Sheorey and A. Lakshminarayan, ``Effect of classical bifurcations on the quantum entanglement of two coupled quartic oscillators'', Phys. Rev. E \textbf{77} (2), 026213 (1-7) (2008).
\bibitem{key-13}N. B. Tufillaro, T. M. Mello, Y. M. Choi and A. M. Albano, ``Period doubling boundaries of a bouncing ball'', J. Phys. France \textbf{47} (9), 1477-1482 (1986).
\bibitem{key-14}N. B. Tufillaro and A. M. Albano, ``Chaotic dynamics of a bouncing ball'', Am. J. Phys. \textbf{54}, 939-944 (1986).
\bibitem{key-15}A. Heck, T. Ellermeijer and E. Kedzierska, ``Striking results with bouncing balls'', In C.P. Constantinou \& N. Papadouris (Eds.), Physics curriculum design, development and validation: Print proceedings GIREP, Nicosia, Cyprus, 190-208 (2008).
\bibitem{key-16}A. Kini, T. L. Vincent and B. Paden, ``The bouncing ball apparatus as an experimental tool'', J. Dyn. Sys. Meas. Control \textbf{128}, 330-340 (2005).
\bibitem{key-17}A. Okninski and B. L. Radziszewski, ``Simple model of bouncing ball dynamics: Displacement of the limiter assumed as a cubic function of time'', Differ. Equ. Dyn. Syst. \textbf{21} (1 \& 2), 165-171 (2013).
\bibitem{key-18}I. Daubechies, \textit{Ten lectures on Wavelets}, Society for industrial and applied mathematics, Philadelphia, Pennsylvania (1992) and references therein.
\bibitem{key-19}A.N. Kolmogorov, ``Dissipation of energy in locally isotropic turbulence'', Pro. R. Soc. Lond. A \textbf{434}, 15-17 (1991). 
\bibitem{key-20}W. Heisenberg, ``On the theory of  statistical and isotropic  turbulence'', Pro. R. Soc. Lond. A \textbf{195} (1042), 402-406 (1948).
\bibitem{key-21}M. Farge, N. Kevlahan, V. Perrier and E. Goirand, ``Wavelets and Turbulence'', Proceedings of the IEEE \textbf{84} (4), 639-669 (1996).
\bibitem{key-22}U. Das, H.S.S. Sinha, S. Sharma, H. Chandra and S.K. Das, ``Fine structure of the low-latitude mesospheric turbulence'', J. Geophy. Res. \textbf{114}, D10111 (1-11) (2009).
\bibitem{key-23}P. Manimaran, P. K. Panigrahi, J. C. Parikh, ``Wavelet analysis and scaling properties of time series'', Phys. Rev. E \textbf{72}, 046210 (1-5) (2005) and references therein.
\bibitem{key-24}J. W. Kantelhardt, S. A. Zschiegner, E. K. Bunde, S. Havlin, Armin Bunde and H. E. Stanley, ``Multifractal detrended fluctuation analysis of non-stationary time series'', Phys. A \textbf{316}, 87-114 (2002).
\bibitem{key-25}K. Briggs, ``Simple experiments in chaotic dynamics'', Am. J. Phys. \textbf{55} (12), 1083-1089 (1987) .
\bibitem{key-26}D. Lai and G. Chen, ``Statistical Analysis of  Lyapunov Exponents from Time Series: A Jacobian Approach'', Math. Comput. Modeling \textbf{27} (7), 1-9 (1998).
\bibitem{key-27}C. K. Peng, S. V. Buldyrev, S. Havlin, M. Simons, H. E. Stanley and A.L. Goldberger, ``Mosaic organization of DNA nucleotides'', Phys. Rev. E \textbf{49} (2), 1685-1689 (1994).
\bibitem{key-28}H. E. Hurst, ``Long-term storage capacity of reservoirs'', Trans. Am. Soc. Civ. Eng. \textbf{116}, 770–808 (1951).
\bibitem{key-29}J. Feder and P. Bak, ``Fractals'', Physics Today \textbf{42} (9), 90 (1989).
\bibitem{key-30}T. Hwa and M. Kardar, ``Dissipative transport in open systems: An investigation of self-organized criticality'', Phys. Rev. Lett. \textbf{62} (16), 1813-1816 (1989). 
\bibitem{key-31}H. J. S. Feder and J. Feder, ``Self-organized criticality in a stick-slip process'', Phys. Rev. Lett. \textbf{66} (20), 2669-2672 (1991); S. Hergarten, \textit{Self-Organized criticality in Earth Systems}, Springer-Verlag Berlin Heidelberg (2002) and references therein.
\bibitem{key-32}R. N. Mantegna and H. E. Stanley, ``Scaling behavior in the dynamics of an economic index'', Nature \textbf{376}, 46 - 49 (1995).
\bibitem{key-33}S. Mallat, \textit{A Wavelet Tour of Signal Processing}, 3rd edition, Academic Press, Elsevier (2009) and references therein.
\bibitem{key-34} C. Torrence and G. P. Compo, ``A Practical Guide to Wavelet Analysis'', Bulletin of the American Meteorological Society \textbf{79} (1), 61 (1998).
\bibitem{key-35}B. B. Mandelbrot, J. W. van Ness, ``Fractional Brownian motions, fractional noises and applications'', SIAM Rev. \textbf{10} (4), 422-437 (1968); B. B. Mandelbrot and N. Goldenfeld, ``Fractals and Scaling in Finance: Discontinuity Concentration Risk'', Physics Today \textbf{51} (8), 59 (1998).
\bibitem{key-36}B.T. Grenfell, O.N. Bjornstad and J.Kappey, ``Travelling waves and spatial hierarchies in measles epidemics'', Nature \textbf{414}, 716-723 (2001).

\end{thebibliography}
\end{document}